# Highly Accurate and Reliable Wireless Network Slicing in 5<sup>th</sup> Generation Networks: A Hybrid Deep Learning Approach

Sulaiman Khan[1], Suleman Khan[2], Yasir Ali[1], Muhammad Khalid[3*], Zahid Ullah[4] and Shahid Mumtaz[5]

[1]Department of Computer Science, University of Swabi, Swabi, Pakistan
[2]Department of Computer and Information Sciences, Northumbria University, UK
[3]Department of Computer Science and Technology, University of Hull, Hull HU6 7RX, UK. (m.khalid@hull.ac.uk)
[4]Department of Computer Science, Institute of Management Sciences, Peshawar, Pakistan.
[5]Instituto De Telecomunicaces, Aveiro, 3750011, Portugal.

*Correspondence: Muhammad Khalid – (m.khalid@hull.ac.uk)

**Abstract** – In current era, the next generation networks like 5<sup>th</sup> generation (5G) and 6<sup>th</sup> generation (6G) networks requires high security, low latency with a high reliable standards and capacity. In these networks, reconfigurable wireless network slicing is considered as one of the key element for 5G and 6G networks. A reconfigurable slicing allows the operators to run various instances of the network using a single infrastructure for better quality of services (QoS). The QoS can be achieved by reconfiguring and optimizing these networks using Artificial intelligence and machine learning algorithms. To develop a smart decision-making mechanism for network management and restricting network slice failures, machine learning-enabled reconfigurable wireless network solutions are required. In this paper, we propose a hybrid deep learning model that consists of convolution neural network (CNN) and long short term memory (LSTM). The CNN performs resource allocation, network reconfiguration, and slice selection while the LSTM is used for statistical information (load balancing, error rate etc.) regarding network slices. The applicability of the proposed model is validated by using multiple unknown devices, slice failure, and overloading conditions. An overall accuracy of 95.17% is achieved by the proposed model that reflects its applicability.

## 1. Introduction

In this modern technological age mobile communication is an important aspect of human lives due to which the communication devices are growing exponentially [1]. These devices require high bandwidth, mobility, low latency, and better quality of service (QoS) to fulfill the needs of future generation communication. Rapid evolution of communication technology from 2G towards 4G and now upcoming 5G and 6G are prominent examples [2]. The future generation communication also require reliability, seamless operations, and reconfiguration management in heterogeneous wireless networks [3]. The service providers are continuously struggling to fulfill the demands of users and provide reliable communication. To achieve these solutions and fulfill requirements of the 5G networks by expanding the LTE networks to provide higher bandwidth, throughput, and better quality of services.

The 5G networks will provide a richer mobility experience in terms of its services, reconfiguration, infrastructure, and large range of operations. It will provide various opportunities for mobilizing a number of application areas like seamless mobility, traffic monitory, healthcare services etc. In the specification of third generation partnership project (3GPP), network slicing is followed as one of the essential component of the 5G network [4]. Through network slicing, the operator will be able to improve the QoS by network reconfiguration, and by providing portion of their network according to the requirements of their customers. Due to the reconfiguration and network slicing, the providers will save much of the resources that are not required by the users at that particular time. The prominent benefits of reconfiguration and network slicing are reducing the latency, high bandwidth, seamless mobility, and better QoS. For example, in the network reconfiguration and slicing of a vehicle-to-infrastructure, passenger will get infotainment services with high bandwidth while car is sending its critical data for autonomous driving. As a result of reconfiguration and network slicing, the operator will use single infrastructure for delivering numerous services to different users with improved QoS [3]. For achieving the QoS and resource efficiency, critical monitoring of the devices as well as their associate traffic is required. In addition to number of challenges, security attacks in heterogeneous wireless environment is also a severe threat for effecting the performance of the network.

Machine learning has shown its potential in multiple domains for taking important decision in critical environments [5]. In the reconfigurable network environments, machine learning will monitor the status of different devices, analyzes the network slices, and a huge amount of data generated during communication for prediction and important decision making. Machine learning can provide network reconfiguration, optimize resources reservation based on their usage, optimized mobile tower operation according to the requirements, optimum decision abilities, and real time performance analysis. The main objectives of the proposed research work are, to develop a machine learning-based reconfigurable wireless network slicing for 5G network using hybrid deep learning model. This model consists of CNN and LSTM. The CNN performs resource allocation, network reconfiguration, and slice selection while the LSTM is used for statistical information regarding network slices. Main contributions of the proposed research work are:

- Accurate slice assignment is the main problem in for the network service providers. Scenario and need-based allocation for a certain IoT device is a big hurdle for the research community and the services providers. A smart mechanism is required to define a criteria to accurately assign the network slices to an unknown device upon request. First contribution among many other contribution of the proposed research work is the accurate allocation of the network slice assignment to all the incoming new traffic requests.
- Load balancing is another critical issue for the service provider as no optimum load balancing results in cross-talks, no on-time connection establishment, and long wait in queue scenarios. These issues not only results in high revenue loss for the companies, but it mostly diverges the users to other network service providers. Accurate load balancing results in efficient utilization of all the available resources. This is considered as a key problem for today's wireless network service providers to overcome. A smart architecture is required to automatically route all the new incoming requests to the master slice instead, to prevent cross talk, long wait in queue for connection establishment, and other problems. Second contribution of this research work is provide an optimum load balancing mechanism in each network slice.
- Slice failure is the condition when a sudden loss of all the established connection occurs. This scenario becomes more severe in case of emergencies (sudden fire/disaster, earthquake, and other critical health problems). This scenario sometimes leads to human life loss. This is a critical issue for 5G/6G networks to overcome. An intelligent mechanism is required to automatically route all the ongoing calls or request the master slice instead, to prevent connection loss or failure of user requests in a certain network slice. Third and most important contribution of this research work the development of an optimal deep learning based model that ensures no slice failure condition.
- To allocate alternate slice (master file) during the failure of slice or during overloading of a slice. In case of overloading conditions (more than 92% usage of network slice in our case) the new incoming network traffic requests will be directly assigned to the master file. While in case of a slice failure the network traffic is directly assigned to other slices to ensure normal operations of these requests. In short master file will act like a backup slice during a slice failure condition or overloading conditions.

Rest of the paper is organized as follows. Section 2 outlines the related work reported in the proposed field. Section 3 presents the role of machine learning techniques in 5G/6G network domain. Section 4 outlines the proposed hybrid model of the paper, while the details the experiments setup and methodology is explained in section 5 of the paper. Section 6 outlines the performance analysis of the proposed model. While the results and discussion are explained in section 7 followed by the conclusion in section 8.

## 2. Related work

$5^{th}$ generation network is an expanded version of the LTE network with wide range of operation and functionality. Compare to the LTE, it is more flexible, scalable to be adopted for diverse use cases. Network slicing in 5G network provides a flexible environment for creating multiple logical networks over a shared physical infrastructure. Artificial intelligence and machine learning techniques are considered as a decision making tools for predicting and making optimum decisions in sliced-based network environment [5-7]. Significant research work is reported in the proposed field such as Oladejo and Falowo have tried to explore 5G network slicing by analyzing the number of users and number of transmitting power using the MVNO's capacity index [8]. The research work presented by Ma et al., [9] used the NFV as well as the SDN 5G core network architecture. Furthermore, Du and Nakao has applied deep learning models to the RAN using application specific radio spectrum scheduling [6]. Abhishek et al., [10] has propose a priority management solution for smart cities where the prioritization of network traffic is managed. Some other approaches suggested in the proposed field are listed in Table 1 below.

**Table 1: List of techniques proposed for accurate network resource allocation purposes**

| S/No | Technique | Description |
|---|---|---|
| 1. | RRC frame and network slicing architecture | Yoo performed analysis of the literature by finding issues in 5G network and provided the information for the selection of network slice, their standardization, and different slice-independents functions [11]. Furthermore, they also proposed RRC frame for slice selection. |
| 2. | NFV and SDN based slicing model | Kurtz et al., [12] has worked on the 5G new radio air interfaces in which the proposed study uses the NFV and SDN work on slicing shows the potential ability to provide service guarantees as well as the dynamic allocation of data rate in the radio air interfaces. |
| 3. | Matrix exponential distribution technique | In this study [13], the matrix exponential distribution is used for representation of handovers for emergency communications and the safety of public. Their work has provided a more accurate decision regarding handovers by the inclusion of different parameters in their decision process. |
| 4. | Virtual networking mechanism | Abhishek et al., [14] proposed a framework that represents network virtualizations with number of multiple providers that necessitates the network slicing in 5th generation network. Another study perform the selection and assignments of virtual networks or its slices and then assign based upon the QCI and the requirement of security associated to the service requested [15]. Addad et al., [4] have proposed a model which provide a cost efficient deployment of network slices. The model allows mobile network operation in allocating underlying resources according to the requirements of its associated users. However for the requirement of multiple services requested from the same device is not been considered in their work. An another study provides mathematical model for providing demand based slice isolation in addition to the guarantee end-to-end delay in 5th generation network slices [16]. Their proposed solution also reduces the chances of distributed network attack such as denial of service in 5th generation core. |
| 5. | Provisional models | A number of provisional models are introduced by different researchers for third party slice assignment [17]. Their study also provides details discussion regarding the isolation properties. A more safe, efficient service-oriented authentication framework which supports fog computing as well as network slicing for the 5th generation enabled IoT services is provided by Ni et al., [18]. In addition to their proposed framework, a mechanism for preserving configured types of slices and the accessing of service types of users. |
| 6. | Manifold models | Thantharate et al., [19] demonstrated three variant models by using MQTT and CoAP protocol. This model provides a generalized mechanisms for OTA delivery of security patches in IoT devices. This model is good in identifying intrusion attacks in 5G network, IoT network for flooding, injecting different kinds of threads and impersonation. |
| 7. | Neural network based auto-encoded architecture | Rezvy et al., [7] proposed a neural network-based model named as deep auto-encoded dense neural network algorithm for network slicing. |
| 8. | SDN and NFV for hyper connected world | Bradai et al., [20] performed a systematic analysis of the available literature to find the gaps in the available solutions and identify future research directions. |
| 9. | Modified dragonfly algorithm | Tripathi et al., [21] proposed an optimal virtual machine placement model for cloud computing. |
| 10. | DDoS attack detection and mitigation | Alhisnawi and Ahmadi developed a framework for detecting and mitigating DDoS attacks in Named Data Networking paradigms [22]. |

From Table 1 it is concluded that a lot of work has been reported for network frame allocation, slice fragmentation, DDoS attacks identification, ensuring security by applying virtualization techniques or other neural network based models, but there is no significant work reported for automatic slice allocation in 5G/6G networks. Nor any work reported for fault detection, load balancing problems, alternate slice allocation in 5G/6G architecture during slice failure or over-flow conditions. To address all these problems a hybrid machine learning based model is proposed in this research work to ensure accurate allocation of network slice based on the incoming new traffic requests, optimum utilization of network resources, load balancing, fault detection, and assignment of master file in case of slice failure or slice over-flow conditions.

## 3. The role of Machine Learning in 5G Network Slicing

Machine learning has shown greater potential in various application domains for solving the decision making problems and perform accurate prediction and classification. Such example includes health care, medical imaging, route selection, data manning, decision support system, resource allocation in network environment [5, 23]. Rapid growth in devices and user demands regarding mobile communication results high speed 3G, 4G, and upcoming $5^{th}$ generation networks. Due to the richer mobility, reliability, and services associated with 5G network, the operators are constantly working in cooperating these feature for attracting more users and to provide better quality of services [3, 24]. Among the prominent feature of 5G technology, network slicing is one of the key element which will allow the operator to customize their capabilities and services according to the need of use cases in networking environment [8, 11]. The network slicing provide more cost effective solution for resource efficiency as well as improving the QoS. 5G network will revolutionise the shape of communication in many areas include media, entertainment, healthcare, social media interaction, networking capabilities, autonomous driving due to its high support for bandwidth and richer set of services [3]. All of these services and huge amount of data during communication will required intelligent decision making supports and predictive decision making for better resources utilization and performance efficiency.

In 5G network slicing, artificial intelligent and machine learning support will pave the ways in taking the automatic decision making and prediction regarding resource assignments, identifying application requirement by monitoring of the network traffic as well as devices status. Machine learning has shown its potential in many applications. Huge amount of data traffic will be involve in 5G mobile communication from large number of devices [24]. Automatic network analysis will be an effective tool for analysing network traffic and make the most optimal decision and prediction to efficiently adjust the services to different slicing. Analysis on big data will provide useful insight for adaptability of network services to avoid network congestion. Deep learning for automation in 5G slicing will results in achieving the task of resources utilization without the intervention of human. Furthermore, automatic monitoring and real-time analysis of data will results improving the slice network performance, detection fraud, security of the communication, and load management. Experts are actively working for exploiting machine learning and deep learning for improving the reliability, efficiency, and performance of network slicing to provide cost effective and improve quality of services to the end users [7, 15, 18]. A typical deep neural network (DNN) model is shown in Figure 1. It consists of an input layer, hidden layer(s) and output layer.

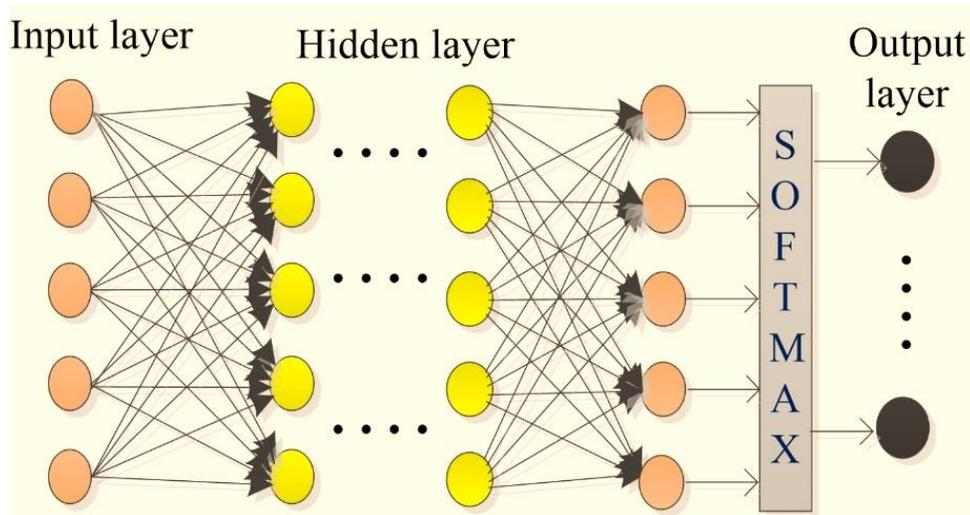
**Figure 1: Typical DNN model**

In the proposed study, a hybrid AI-based solution is provided for optimizing efficiency of 5G network slicing in the context of load balancing, accurate resource allocation and restriction of network slice failure and alternate assignment of devices involved in the communication process. Furthermore, researchers are actively working in providing AI and Deep learning based solutions for improving efficiency of 5G network slicing.

## 4. The Proposed Hybrid Model

5G network slicing is a challenging task in future generation wireless networks, commercial businesses and mobile operators. It followed as one of the key element for 5G technology. It allows the mobile operators to simulate multiple instances of the network using a single base for better quality of services. In order to develop a smart decision making mechanism for incoming network traffic, to ensure load balancing, restricting network slice failure, and provide alternate slice in a case of slice failure or overloading conditions, machine learning based reconfigurable wireless network solution is required. To address these key problems and present a model to efficiently predict unknown network requests, perform load balancing, optimum utilization of resources, and restricting network slice failure a hybrid deep learning based model is proposed in this research work that consists of CNN and LSTM models as depicted in Figure 2. Due to high capabilities and no data-driven nature of hybrid models, these models are selected in many research problems such as: traffic flow preserving purposes [25], efficient resource allocation in IPv6 [26], and many others.

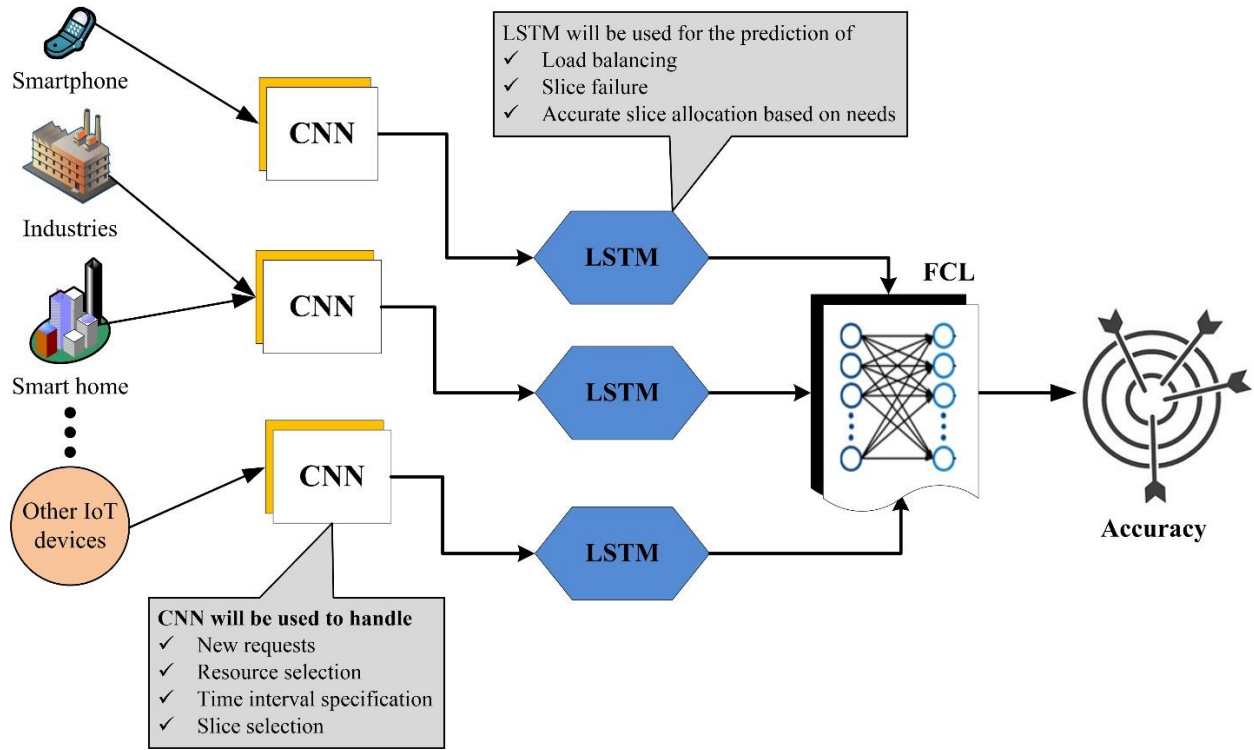

**Figure 2. Proposed hybrid model**

In Figure 2 the CNN architecture is responsible for the detection of network traffic, allocation of network slice and accurate resource assignment. Generally, the LSTM networks are good for the prediction-oriented tasks. So, keeping in view the capabilities of the LSTM classifier it is used for the statistical performance like; load balancing and assignment of alternate slice (master file) in the situations such as, network slice failure or accurate assignment of network resources. This hybrid model outperformed by generating high accuracy rates, low miss-classification error rates, and comparatively small experimental and simulation time. Also this model generated high capabilities for small amount of data and ultimately we can say that the proposed model is not a data hunger model.

After testing for various conditions of: varying training and test sets, time consumption, unknown devices requests, slice failure, slice overloading, imbalanced allocation of resources and accuracy. The proposed model outperformed by generating an accuracy rate of 95.17% that reflects the applicability of the proposed model for the selected research problem.

Algorithm for the proposed hybrid model is given below.

| **Algorithm: To model 5G network slicing** |
|---|
| **Input:** Different device types (smart phones, smart homes, automotive, industrial devices, and so on) <br> **Output:** Network slicing, load balancing, no network slice failure, alternate network slice for incoming network traffic in case of slice failure |
| **Start** <br> *Step 1:* initialize eMBB, mMTC, URLLC, mFile as an empty vector of length k     // mFile represents "master file" <br> *Step 2:* initialize slice = 0                              // to ensure operations of a certain network slice <br> *Step 2:* initiate network request based on different input types, req |

```
    While i <=k

// to control load balancing in different network slices
        s1 = eMBB_k/sizeof(eMBB) × 100    // s1 stores the percentage size utilized by eMBB slice
        s2 = mMTC_k/sizeof(mMTC) × 100    // s2 stores the percentage size utilized by mMTC slice
        s3 = URLLC_k/sizeof(URLLC) × 100  // s3 stores the percentage size utilized by URLLC slice
        s4 = mFile_k/sizeof(mFile) × 100  // s4 stores the percentage size utilized by mFile slice

    if (function)                         // if the slice works properly (no failure and no overloading)
        if (high throughput and s1 <=92%) // utilization above 92% is followed as overloading in our case
            eMBB_k = req_i                // in case of high throughput device should be assigned eMBB slice
        ElseIf ((reliability && low latency) and s2 <=92%)
            mMTC_k = req_i
        ElseIf ((low throughput && high density) and s3 <=92%)
            URLLC_k = req_i
        Else
            mFile_k = req_i               // if utilization size exceeds 92% then master file resources must
                                          // be allocated to the incoming network traffic
    Else
        mFile_k = req_i                   // if a certain network slice fails then the requests are automatically
                                          // assigned to master file
    End if

  End while
```

- In this algorithm "i" and "k" represents empty arrays that are used to keep the records (number of users) of each network slice. In our case "k" is used to measure the statistical information (percentage utilization) of a certain slice (eMBB slice, URLLC slice, mMTC slice) while "i" counts the number of new incoming network traffic requests.
- Since eMBB devices requires a real-time and need base connection. For exampling in case of any miss-happen or emergencies we need high throughput and real time connection. That's why we put an "*if statement*" with high throughput and s1<= 92% (in our case 92% is upper bound for a particular slice utilization).
- As URLLC requires high bandwidth for data transmission (videos). Normally low throughput but high density is required. So, we put a conditional statement that if "***high density and low throughput***" is required then all the incoming new traffic will be assigned to URLLC slice.
- While if incoming request needs a reliable and low latency connection. Then all the incoming traffic must be automatically assigned to the Mmtc slice.

## 5. Research Methodology

Neural network and machine learning (ML) techniques play a significant role in many industrial applications such as; internet security [27, 28], text recognition [29, 30], security purposes [31], wireless localization for indoor navigation purposes [32], and many others. Due to its automatic feature extraction capabilities and high recognition rates the neural networks will be applied to the IoT devices, as these devices generates a massive data over the 5G network. Accurate analysis and quick decision regarding a particular input device is too difficult for human. To address this problem a machine learning-based hybrid model is proposed in Figure 3 that is capable of deciding which network slice is suitable for a new connected device. The proposed hybrid model also aims to detect network load, slice failure and the decision for the adjustment of a particular network slice for the newly connected unknown device. For calculating and statistical purposes long short term memory is used in this research work while for deep slice purposes an artificial neural network based on convolution neural network is proposed. The proposed model is trained using the same dataset [33] consist of 65000 unique entries.

This dataset consists of the key performance indicators (KPIs) regarding network and devices. It consists of type of devices (IoT-based devices, smartphones, URLLC devices, and many others) connected to the internet, user equipment (UE) category, maximum packet loss, QoS class identifier (QCI), packet delay budget, day and time information, whether conditions (normal condition or harsh conditions), and many others. As depicted in Figure 3 multiple devices including healthcare to personal use devices, smart phones to vehicle systems, athletes performance measure to physical performance measures and so on request for one or multiple sources. There are multiple UE category values assigned to each device and a pre-defined QCI values assigned upon incoming new reequest. In 5G network, both packet loss rate and packet delay budget are integral elements of the 5QI where 5QI is 5G QoS Identifier. The proposed hybrid model will also keep weekly information for the requests receivable in the communication system. This overall information will be shuffled and processed by the proposed deep learning based hybrid model to smartly decide the allocation of the available resources and predict accurately the reservation of the network resource for the new incoming traffic in future.

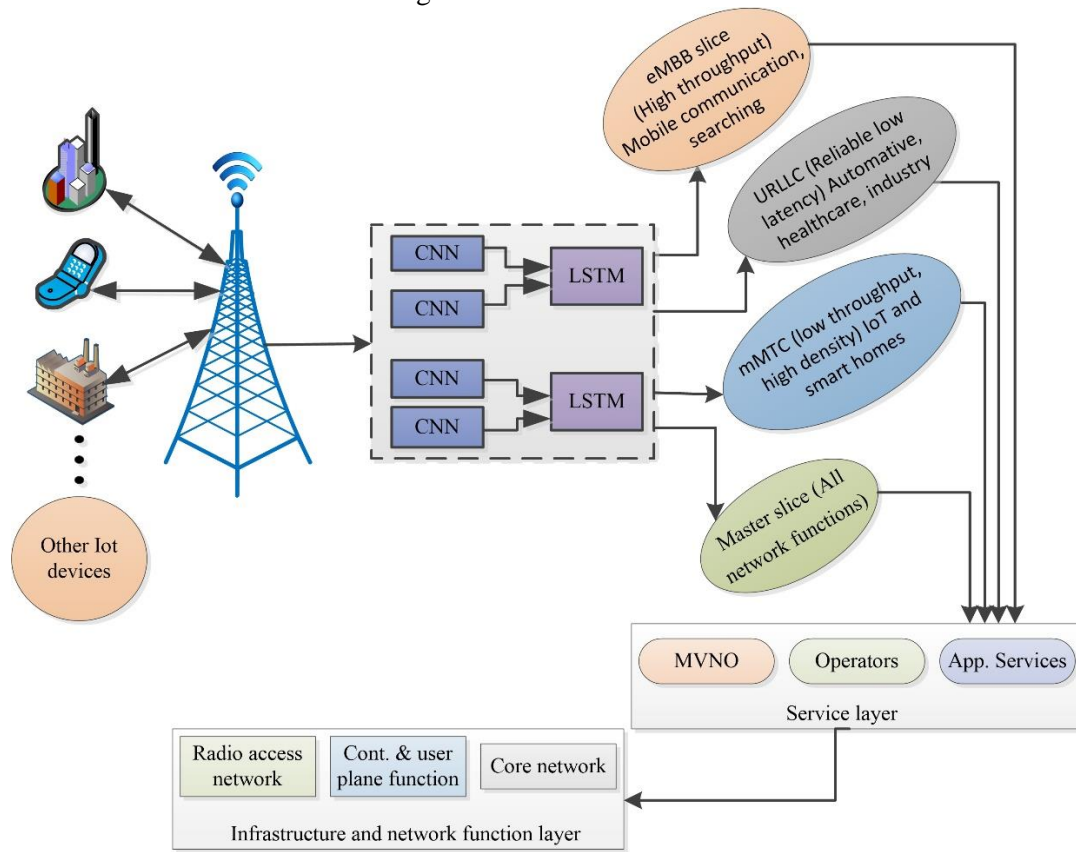

**Figure 3: Proposed methodology**

In this research work different parameters are selected for the development of the CNN architecture. These parameters are detailed in Table 2.

**Table 2. Parameters of the CNN and LSTM model**

| S. No | Parameter | Value |
|---|---|---|
| 1. | Total number of layers | 7 |
| 2. | Number of hidden layers | 5 |
| 3. | Activation function | relu |
| 4. | Performance metrics | Accuracy, Time, F-measure, Specificity, True-False values, varying training and test sets |

All these steps are discussed in details below.

## 6. Performance Evaluation

A simulating model is developed in python using Tensor-Flow and Keras libraries to test the performance of the proposed model. These libraries are the best practising tools for the neural network based architectures development purposes. While NS2 is used for the simulation of the incoming network traffic. Mesh UAV network topology is used in this research work. As it has a good performance in terms of reliability and flexibility. This is an ideal network topology when fast response is required in hard and difficult conditions. Performance of multiple connected devices is tested based on request duration, packet loss and delay budget rates, and predicted slice. Overall performance results are depicted in Table 3. During the simulation phase some variations are further sub-classified such as mMTC devices are further divided into two broad classes (1) mMTC class that requires a real time connection for data sending and receiving purposes and (2) the mMTC class that required a need-based connection for data transferring purposes. Smartphones are mostly used by common people for calling, web-browsing, but at the same time if a person wants to respond in an emergency situation like any miss-happen then the model must ensure lower packet loss and packet delay. The slice categories consists of enhanced Mobile Broad Band (eMBB), massive Machine Type Communication (mMTC), Ultra Reliable Low Latency Communication (URLLC), and the Master slice. The Master slice has the network functionalities of all the corresponding network slices. It act like a back-up slice and provides resources to other slices based on hard load situations on other slices.

**Table 3. Performance evaluation of proposed hybrid slicing model**

| S.No | Devices | Packet loss rate | Packet delay budget (ms) | Duration (sec) | Slice identified |
|---|---|---|---|---|---|
| 1. | Healthcare | $10^{-6}$ | 15 | 200 | URLLC |
| 2. | Intelligent transportation | $10^{-6}$ | 15 | 50 | URLLC |
| 3. | Smart cities | $10^{-3}$ | 60/300 | 90 | mMTC |
| 4. | IoT devices | $10^{-3}$ | 60/300 | 50 | mMTC |
| 5. | Smart phones | $10^{-3}/10^{-6}$ | 50/75/100/130/300 | 250 | eMBB |
| 6. | Industry 4.0 | $10^{-3}/10^{-6}$ | 15/50 | 160 | mMTC/ URLLC |
| 7. | Unknown devices | $10^{-3}/10^{-6}$ | 15/50/60/75/110/150/300 | 40/110/190 | eMBB/ mMTC/ URLLC |

In the proposed hybrid model, the network traffic is predicted using the prior information in an every individual network slice. This information (overall network connections) will help to identify which keep 'network slice' is being utilized the most and to keep track of the connections. This will ultimately results in optimally distributing the incoming traffic requests between all the slices as required. DNN architecture are handy in such cases; as there is no clear set of rules how to treat all the incoming network traffic (network devices). Cellular handovers, on the other hand based upon different network elements. With each new incoming request, an intelligent model is required to automatically learn and adapt very quickly to meet the changes or the new requirements of the incoming request. DNN can help in such cases to accurately identify and accommodate the unknowns in the network.

For training the proposed model 65% of the data is used for the training purposes while the remaining 35% of the data is used for testing purposes. Figure 4 depicts the simulation results of hybrid deep learning model executed for the first 20 hours. After completion it results in providing the number of users being served at a particular interval of time. To test the applicability of the proposed model the model is tested consecutively for three days for 24 hours and it outperformed by generating efficient results in terms of assigning alternate slice (master file) in case of an individual slice failure or over-flow conditions. 92% utilization of a certain network slice is selected as upper bound in this research work. So if the number of requests exceeds this upper bound limit then alternate slice is automatically assigned by the proposed hybrid model to ensure reliable connection, high throughput, high security, automatic allocation of most relevant network slice, and required optimum bandwidth size for each request.

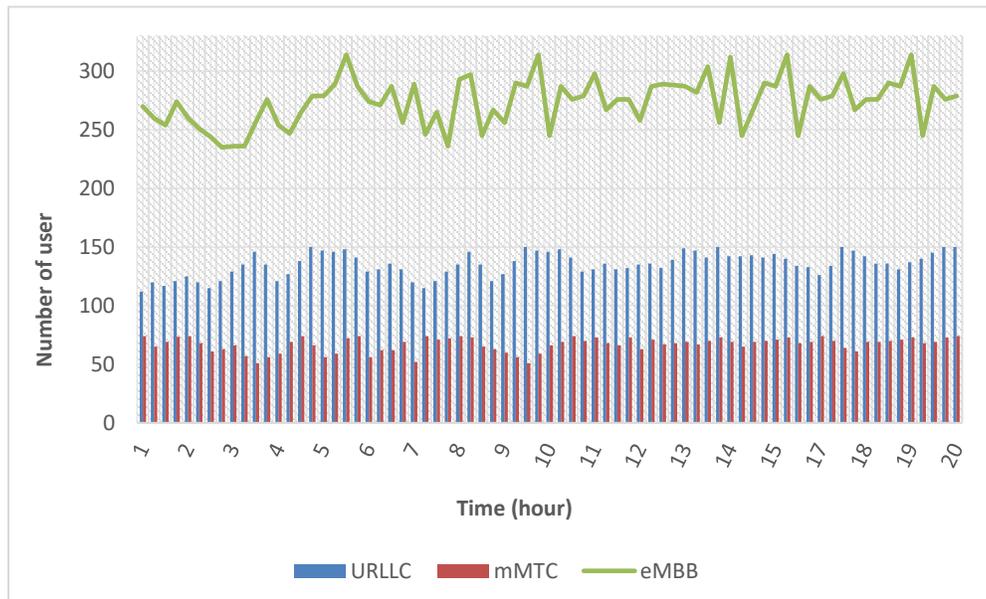

**Figure 4. Number of active user observed after every 10 minutes**

In a 20-hour simulation, approximately half million user connection requests were generated that consists of 45% for eMBB, 20% for mMTC and 35% for URLLC. For the proposed model the plot begins when the model reached a regular state, in Figure 4 it is marked as 1st hour. As depicted in Table 3, all incoming new traffic has a selected time-to-live (TTL). For example, in eMBB slice there were 224 active users at a particular interval of time. eMBB is allotted more TTL values than URLLC and mMTC because of high broadband and multi-active users. This will ultimately help in analyzing the user patterns and made optimum decision capabilities using the information accumulated from the connected network device.

Hybrid deep learning model will finally decide which device should be allocated to which network slice and will also will evaluate all the incoming future connections for accurate network slice allocation. It will ultimately help the network model to prepare for any new incoming connections to be properly allocated by assigning optimum resources. The dataset also includes the day and timing information for new connections. And it also be aware of the network slices to be assigned for a particular network connection using the prior information of the devices.

## 7. Results and Discussions

This section of the paper outlines the performance analysis of the proposed hybrid model in terms of load balancing, slice prediction and network availability. To test the applicability of the proposed model three different scenarios are formulated in this research work:

- **Accurate slice assignment** – Accurate slice assignment is the main problem in for the network service providers. Scenario and need-based allocation for a certain IoT device is a big hurdle for the research community and the services providers. A smart mechanism is required to define a criteria to accurately assign the network slices to an unknown device upon request.
- **Load balancing** – Load balancing is another critical issue for the service provider as no optimum load balancing results in cross-talks, no on-time connection establishment, and long wait in queue scenarios. These issues not only results in high revenue loss for the companies, but it mostly diverges the users to other network service providers. Accurate load balancing results in efficient utilization of all the available resources. This is considered as a key problem for today's wireless network service providers to overcome. A smart architecture is required to automatically route all the new incoming requests to the master slice instead, to prevent cross talk, long wait in queue for connection establishment, and other problems.
- **Slice failure scenario** – Slice failure is the condition when a sudden loss of all the established connection occurs. This scenario becomes more severe in case of emergencies (sudden fire/disaster, earthquake, and other critical health problems). This scenario sometimes leads to human life loss. This is a critical issue for 5G/6G networks to overcome. An intelligent mechanism is required to automatically route all the ongoing calls or request the master slice instead, to prevent connection loss or failure of user requests in a certain network slice.

The hybrid model will capture the time and other relevant information regarding slice failure in a certain situation that will ultimately makes the model smart in deciding before assigning a slice to the requested device types. Also it helps in advance preparing for the proposed model.

These scenarios are discussed in details below to validate the proposed model.

### 7.1. Accurate slice assignment

The proposed hybrid deep learning model is trained for multiple input types (smart phones, smart cities, healthcare devices, and so on) based on network and device KPIs. An ovrall accuracy rate of 95.17% is calculated as depicted in Figure 8. The model is also tested for unknown input types with random parameters. Slice prediction accuracy was 96.46% for unknown devices. Table 4 shows 5 unknowns devices and the corresponding information regarding network slice assigned, packet loss measured, technology used, and packet budget rate recorded.

**Table 4. Slice prediction and assignment for unknown devices**

| S.No | Device type | Devices/technology | Packet loss rate | Packet delay budget | Predicted slice |
|---|---|---|---|---|---|
| 1. | Unknown device – 1 | IoT or LTE/5G | $10^{-2}$ | 100 | mMTC |
| 2. | Unknown device – 2 | LTE/5G | $10^{-6}$ | 10 | URLLC |
| 3. | Unknown device – 3 | LTE/5G | $10^{-2}$ | 60 | eMBB/mMTC |
| 4. | Unknown device – 4 | IoT | $10^{-6}$ | 150 | eMBB |
| 5. | Unknown device – 5 | IoT | $10^{-3}$ | 180 | eMBB |

### 7.2. Load balancing

Applicability of the proposed hybrid model is also validated for overloading scenarios. If the number of connections exceeds a certain threshold values, in our case we select 92% utilization of the slice.

Figure 5 shows that an mMTC slice is detected to have over 92% utilization (the question mark represents the over-utilization problem of the mMTC slice). In other words it exceeded the defined threshold value, so, the master slice acts as backup for any new mMTC connections. Our hybrid model can accurately identify this overloaded scenario and can automatically redirect the newly network traffic to the next slice without causing overloading scenarios in the slices. After comparing the overloading condition in Figure 5, the master slice assigns the required resources to the overloaded new connections as depicted in Figure 6.

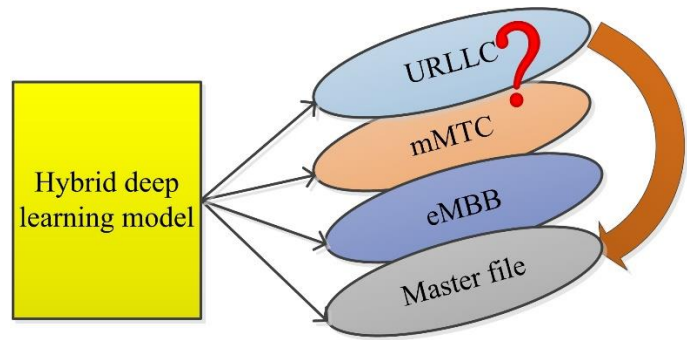

**Figure 5. Load balancing model for slice overloading conditions**

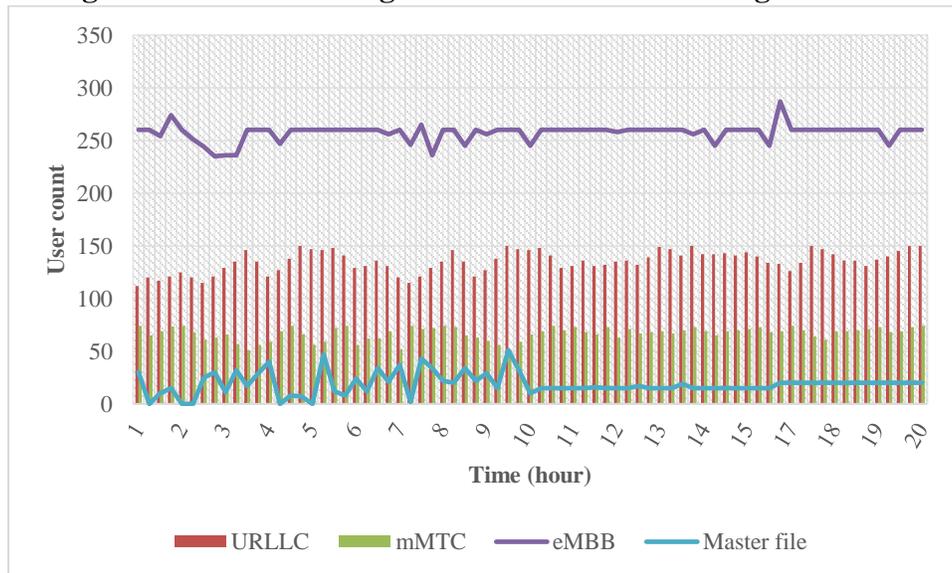

**Figure 6. Load balancing when connection requests exceeds a certain limit**

### 7.3. Slice failure conditions

To check the applicability of the proposed hybrid model a complete slice failure condition was generated as depicted in Figure 7, and the model is tested. To handle this scenario the proposed model routed all new URLLC related traffic to the master slice. This automatic allocation helped in avoiding the traffic loss at the network side. But due to sudden failure, any ongoing communication on that particular slice would be impacted and the connections can be lost. It was predicted during the simulations that, to ensure no connection loss and optimum transmission then care must be taken before such conditions arised.

Figure 8 depicts that our simulated model had failures on the mMTC slice for a period of three and half hours and fifteen minutes from 2:30hr to 4:45hr on the mMTC slice for one period and for another four-hour period 13hr to 17hr (The question mark represents the failure condition of the mMTC slice for a specific interval). The master slice acts like a backup and accurately route all the network traffic during these failure conditions.

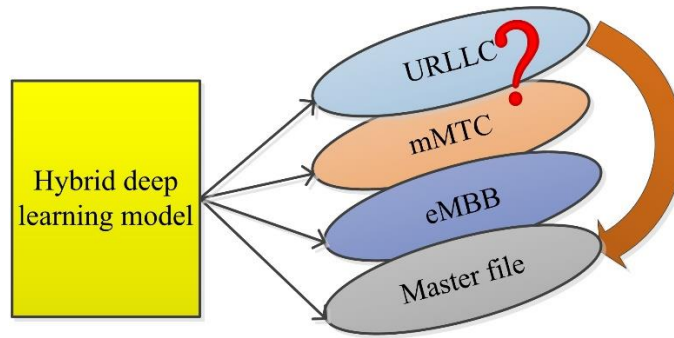

**Figure 7. mMTC slice failure and allocation to master file to overcome failure conditions**

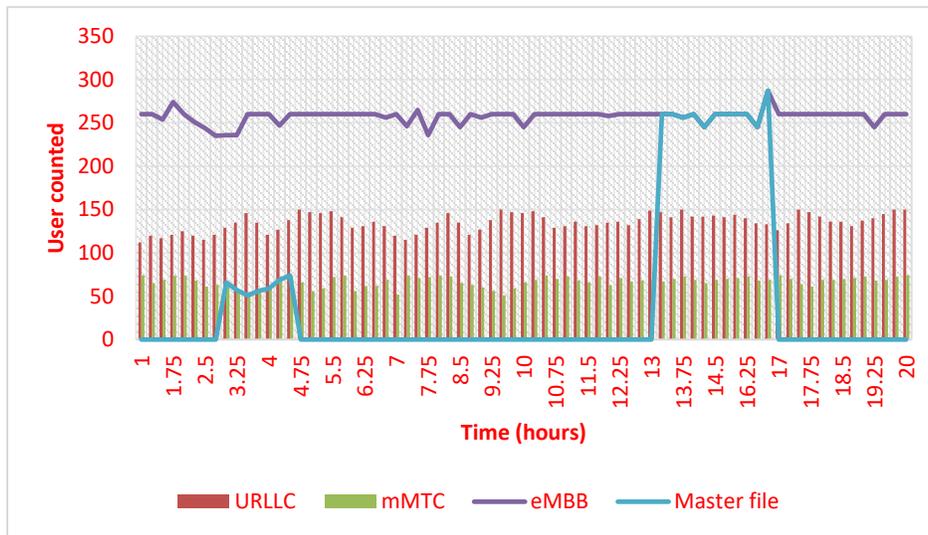

**Figure 8. URLLC slice failure and allocation to master slice**

The model is also tested for the whole week for both day and night timings based on varying training and test sets, and time information to ensure the applicability of the model in both of these conditions. An overall accuracy rate of 95.17% is achieved for the proposed model as depicted in Figure 9. This high accuracy rate of the proposed hybrid reflects the applicability of the proposed model in 5G network slicing.

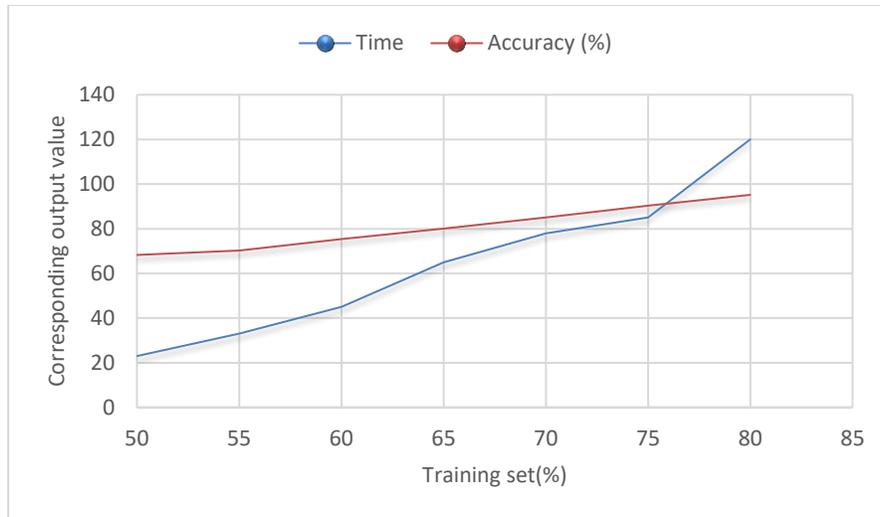

**Figure 9. Hybrid model accuracy results**

The proposed hybrid model based on different performance metrics of accuracy, recall, precision, F-score and misclassification rate it was concluded that the model outperforms by generating a recognition rate of 95.17% for the network slice. The results generated are shown in Table 5.

**Table 5: Proposed model efficiency based on difference performance metrics**

|  | Performance metrics | | | |
| --- | --- | --- | --- | --- |
|  | *Accuracy* | *Recall* | *Precision* | *F score* |
| **Output generated (%)** | 95.17 | 95.54 | 95.76 | 93.59 |

## 8. Conclusion

5G network slicing is a challenging task and considered as one of a significant feature for next generation wireless networks and commercial businesses. The development of a smart decision making system for incoming network traffic to ensure load balancing, restricting network slice failure, and provide alternate slice in a case of slice failure or overloading conditions is a big hurdle for the research community. To address this problem, the proposed research work outlines the benefits of using hybrid slicing mechanism for optimum prediction of the, best appropriate network slice for all the incoming network traffic based on device key features. This hybrid model is capable of handling different key issues in 5G networks such as, network slice failure and load balancing. Both of these issues are severe for any network service provider. As a certain slice failure results in the form of connection loss for each ongoing calls or newly established requests. While load balancing is another critical issue for the service provider as no optimum load balancing results in cross-talks, no on-time connection establishment, and long wait in queue scenarios. These issues not only results in high revenue loss for the companies, but it mostly diverges the users to other network service providers. This model ensures no connection loss and optimum load balancing conditions by routing both the ongoing requests (in case of slice failure) and the new incoming requests (in case of over-flow slice conditions) to master slice. The capabilities of the model are also tested using other performance metrics such as specificity, recall, time consumption, varying training and test sets, true-false rates and f-score. An overall recognition rate of 95.17% is reported for the proposed hybrid model that reflects the applicability of the proposed approach

In future we want to test the applicability of the proposed model in a real production environment once the 5G ecosystem commercially available for the consumers along with devices and networks facilities. Also we will further enhance this research work to significantly address the issues of: handovers, caching, predicting the future load and accordingly assignment of resources, application-based slice management and borrowing network resources from other slices.

**Conflict of interest**
The authors have no conflict of interest regarding the paper.